\documentclass[journal]{IEEEtran}

\usepackage[compress]{cite}
\usepackage{graphicx}
\usepackage{amsthm, amsmath}
\usepackage{url}
\usepackage{subcaption}
\usepackage{algorithm}
\usepackage{algpseudocode}
\usepackage{algcompatible}
\usepackage[super]{nth}
\usepackage[english]{babel}
\usepackage{makecell}
\usepackage{multirow}
\usepackage{color}

\begin{document}

\title{Opportunistic Selection of Vehicular Data Brokers as Relay Nodes to the Cloud} 

\author{Shadha Tabatabai~\IEEEmembership{Student Member,~IEEE}, Ihab Mohammed~\IEEEmembership{Student Member,~IEEE}, Ala Al-Fuqaha~\IEEEmembership{Senior Member,~IEEE}, and Junaid Qadir~\IEEEmembership{Senior Member,~IEEE}}

\maketitle

\begin{abstract}

The Internet of Things (IoT) revolution and the development of smart communities have resulted in increased demand for bandwidth due to the rise in network traffic. Instead of investing in expensive communications infrastructure, some researchers have proposed leveraging Vehicular Ad-Hoc Networks (VANETs) as the data communications infrastructure. However VANETs are not cheap since they require the deployment of expensive Road Side Units (RSU)s across smart communities. In this research, we propose an infrastructure-less system that opportunistically utilizes vehicles to serve as \textit{Local Community Brokers} (LCBs) that effectively substitute RSUs for managing communications between smart devices and the cloud in support of smart community applications. We propose an opportunistic algorithm that strives to select vehicles in order to maximize the LCBs' service time. The proposed opportunistic algorithm utilizes an ensemble of online selection algorithms by running all of them together in passive mode and selecting the one that has performed the best in recent history. We evaluate our proposed algorithm using a dataset comprising real taxi traces from the city of Shanghai in China and compare our algorithm against a baseline of 9 Threshold Based Online (TBO) algorithms. A number of experiments are conducted and our results indicate that the proposed algorithm achieves up to 87\% more service time with up to 10\% fewer vehicle selections compared to the best-performing existing TBO online algorithm.
\end{abstract}

\IEEEpeerreviewmaketitle

\section{Introduction}

The widespread use of smart devices has led to the revolution of the Internet of Things (IoT). According to Gartner, 20 billion devices (things) will be connected by 2020~\cite{gartner_2017}. Many applications in the cloud use data generated by billions of devices for data analysis and decision making in different business sectors such as transportation, aviation, health care, and social networking~\cite{fog_vehicular_2017}. Nevertheless, those billions of devices generate a huge amount of data that is increasing at an exponential rate\footnote{As per one projection, data traffic per subscriber is increasing roughly at a rate of 50 percent annually~\cite{millimeter_2017}.}, which introduces communication and processing challenges~\cite{vehicular_fog_2016}. 

To respond to these challenges, some researchers propose using Vehicular Ad-Hoc Networks (VANETs) to reduce the burden on the cloud by locally using the vehicles for communication as well as computation. VANETs utilize vehicles as computation and communication units to provide services to communities since vehicles travel across communities. Besides, vehicles are equipped with On Board Unit (OBU), which is a device highly capable of processing and communication. In fact, it is expected that nearly 90 percent of vehicles by the year 2020~\cite{OBU_2017} will be equipped with an OBU. 

Additionally, VANETs utilize a number of Road Side Units (RSU), which are road-side devices with computation and communication capabilities. An RSU can use 5G communication technologies to increase data rates.
In spite of that, 5G technologies require dense deployment in small cell formation and suffer from great loss in penetrating buildings and obstacle blockages~\cite{millimeter_2017}. The deployment of RSUs also incurs additional cost and requires maintenance. As a result, this solution is not appropriate for communities with limited communications infrastructure. Furthermore, 5G technologies and their backhaul transport technologies are expensive~\cite{cost_5g_2018}.

To solve this problem, we envision a system where a community is divided into a number of zones.
In each zone, we propose to use one vehicle to serve as a Local Community Broker (LCB). LCB uses the publish-subscribe paradigm to manage the communications between the smart devices locally in the zone and those between smart devices across different zones through the cloud. A smart device can subscribe for a service provided by another smart device located in the same zone or in a different zone through the LCB.
At any given time, one vehicle in each zone (if any exists) is selected to work as an LCB. The selection is made by the Smart Community Management Center (SCMC), which is hosted on the cloud.

In our proposed system, the SCMC represents a Cloud Service Provider (CSP), the LCB is a Cloud Service Broker (CSB), and smart devices are the Cloud Service Customers (CSC). A CSB plays the role of an intermediary between the CSP and the CSC, which has possible benefits from economic and security perspectives~\cite{cloud_brokering_2015}.

Selecting the best LCB is a challenge. When a new vehicle enters a zone, this vehicle is either selected or rejected by the SCMC to serve as an LCB. When a vehicle is rejected, it can not be selected later in the same zone. As a result, the SCMC may regret its decision of rejecting a vehicle especially when new coming vehicles have shorter service times. This problem is of an online nature and is similar to the online hiring problem where once an employee is hired or rejected for a job, the decision can not be reversed in the future (see Section~\ref{onlineHiringAlgorithms}).

While maximizing the total number of hired vehicles to serve as LCBs during the running period increases the service time in a zone (which is desirable since it provides the longest possible coverage possible), it also results in greater vehicle switching (which is costly since some incentive has to be used to convince vehicles to serve as LCBs while in the zone). Besides, subscribers need to be moved from the old LCB to the new one, which results in a delay. Consequently, the more the LCBs used, the more the cost. However, reducing the number of LCBs in a bid to reduce the cost minimizes the service time in the zone. Therefore, we propose an algorithm that utilizes an ensemble of Threshold Based Online (TBO) algorithms for LCB selection with the goal of maximizing service time while maintaining a reasonable number of vehicle switching. 

The proposed algorithm opportunistically selects the best performing TBO online algorithm from its ensemble based on the observed performance in terms of service time in the recent history. The proposed algorithm chooses one algorithm from its ensemble to be active and set all other algorithms to be passive. By doing so, the active algorithm alone selects the vehicle to be used as an LCB while performance of passive algorithms is used to choose the best performing algorithm in the ensemble by choosing the one that demonstrated the highest average service time in the recent history (i.e., greedy approach). Consequently, different TBO algorithms might be utilized over time.

To the best of our knowledge, this is the first research effort that utilizes TBO algorithms for the selection of vehicles to serve as LCBs in support of smart community applications. Even though this work utilizes the proposed algorithm presented in our previous paper~\cite{previous_paper}, the application domain is different in which the previous work focuses on minimizing the total delivery delay of messages using vehicles between the two zones while this work is focused on maximizing the vehicles' service time in a given zone. Furthermore, this work has significant results and insights in this particular vertical application domain of using vehicles as data brokers when contrasted to our previous work that focuses on vehicular data ferrying.

\section{Motivation}

\begin{figure}[!th]
\centering
\begin{subfigure}[b]{0.40\textwidth}\includegraphics[width=\textwidth]{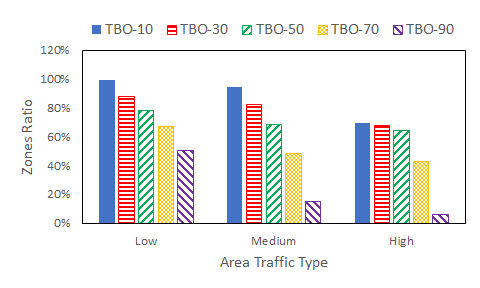}
\caption{Performance of algorithms across zones in terms of service time. \textit{No single algorithm performs the best in all zones in area of high traffic volume.}}
\label{fig:zonesPerAlg}
\end{subfigure}

\begin{subfigure}[b]{0.40\textwidth}\includegraphics[width=\textwidth]{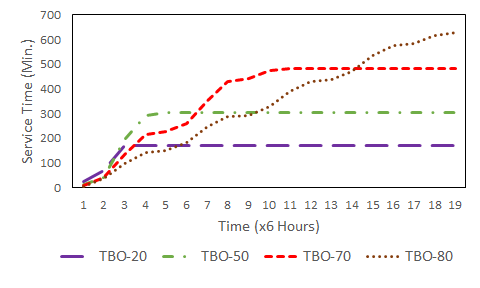}
\caption{Performance of algorithms in one zone over time in terms of service time. \textit{No single algorithm performs the best over time.}}
\label{fig:overTime_noProposed}
\end{subfigure}

\caption{Observations on the performance of algorithms on a single and multiple zones.}
\label{fig:motivation}
\vspace{-5mm}
\end{figure}

We studied the online hiring algorithms and found that some online hiring algorithm can be replaced by a TBO algorithm (see section~\ref{onlineHiringAlgorithms}) with a specific value. Therefore, we analyze the performance of the TBO algorithm in terms of service time using a variety of threshold values that range from low to high values. We found that the TBO algorithm performs better using some of these threshold values in one zone and worse in other zones. To understand this behavior in more depth, nine TBO algorithms are run (explained later in Section~\ref{experiment_results}) with each having a different threshold value. For each of these algorithms, we count the number of zones where the algorithm performance is the best compared with the other algorithms. We found that algorithms perform the best in some but not all of the zones in area of high traffic volume as indicated in Fig.~\ref{fig:zonesPerAlg}. Also, this figure shows that more than one algorithm performed the best in the same zone---in other words, in some zones, a number of algorithms may perform equally well. Moreover, focusing on one zone, we observe that for every period of time, one algorithm perform the best as indicated in Fig.~\ref{fig:overTime_noProposed}. For that reason, we were inspired to develop an intelligent algorithm that studies the history of these nine TBO algorithms and switches to the algorithm that performs best in recent history. By doing so, the intelligent algorithm performs the best over time and across all zones.

\section{Related Work}
\label{related_work}

In recent years, there has been an increasing amount of literature on the use vehicles as a communications infrastructure for smart communities. In this section, We will initially outline the contributions of these existing works, and then we will describe how these existing approaches relate to our work and problem setting.

Alahmadi et al. \cite{Distributed_2017} created a vehicular cloud network model in which a number of vehicles close to a traffic light form a non-permanent vehicular cloud by gathering the clustered computational resources. The researchers aim to reduce the processing and network power utilized in the data center.

Aloqaily et al. \cite{smart_city_2017} introduced a method in which vehicular services in smart cities are provided continuously by using the concept of smart vehicle as a services (SVaaS). Such services include: sensing, storing, computing, infotainment, and/or mobilizing. To provide continuous services, the researchers predicted the location of vehicles in order to prepare the demanded services ahead of time.

Hou et al. \cite{vehicular_fog_2016} state that solutions to computation and communication challenges in vehicular applications like RSU, cellular network, and mobile cloud computing are not applicable due to dependency on existing infrastructure and high cost. The researchers proposed an architecture named Vehicular Fog Computing (VFC) that utilizes a number of collaborating vehicles to perform computation and communication tasks. In another related work~\cite{fog_vehicular_2017}, authors indicate that in some cases, such as peak hours, the computation capacity of fog is overloaded by an increasing number of requests. The authors propose to use VFC for supporting fog computing with more storage and computation power and focused on the use case of parked cars.

Want et al.~\cite{offloading_iov_2018} propose using Internet of Vehicles (IoV) as a fog consisting of a number of vehicles close to RSU for executing real-time traffic management in order to reduce the average response time per vehicle's report. The authors formulate and solve the offloading operation as an optimization problem.

Cao et al.~\cite{qoe_2017} claim that cloud-based solutions have many issues such as network bandwidth bottlenecks, high delay, and low Quality of Experience (QoE). They investigate the use of IoV in edge computing and proposed a strategy for users to pick vehicles to achieve maximum QoE.

In~\cite{adaptive_2017}, the authors claim that some services such as those related to information about local weather and traffic depend on time and geographical location and thus not readily available through the Internet. To make such information available, the authors proposed using a VANET without the dependency on existing infrastructure for retaining this kind of information. Also, they developed a method in which data transmission probability is determined by vehicles depending on the density of data retention in near vehicles.

All the aforementioned works have one or more of the following failings (which makes them ill-suited for our problem setting):

\begin{enumerate}

\item they depend on clusters of vehicles near a traffic light; 

\item they depend on parked vehicles; 

\item they provide specific services to vehicles and not to the cloud in general; 

\item they depend on users for picking service vehicle. Other infrastructure based projects are also not appealing due to the high cost of commissioning the infrastructure. 

\end{enumerate}

In addition, most of the existing works do not approach the problem from an online perspective. Our work is distinctive in that our proposal assumes no infrastructural support and works in an online fashion to hire a vehicle as a local community broker (LCB) regardless of geographical constraints.

\section{System Model}

In this paper, we assume that for every vehicle, the time spent in a zone is estimated by the SCMC. Our assumption is based on many research efforts that appeared in the recent literature 
\cite{assumption_1, assumption_2, assumption_3} to predict this parameter.

\begin{figure}[h]
\centering
\includegraphics[width=.40\textwidth,trim={0 0 0 0},clip]{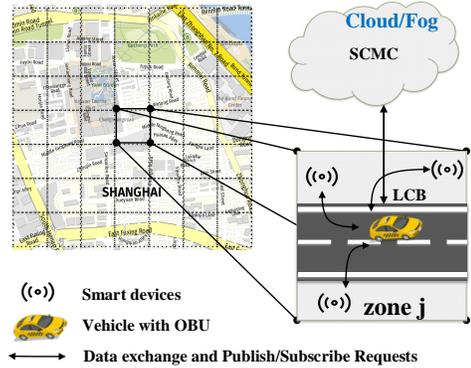}
\caption{Publish-subscribe model between the smart devices and the SCMC through the LCBs.}
\label{fig:SystemModel}
\end{figure}

In our proposed system, to use vehicles as LCBs in a given area, we divide the area of interest (i.e., city) into $Z$ zones. A number of smart devices (i.e., sensors and actuators) exists in every zone. Smart devices subscribe for services provided by other smart devices in the same zone or in a different zone through the LCB. The LCB manages publish/subscribe requests between the smart devices through the SCMC in the cloud as illustrated in Fig.~\ref{fig:SystemModel}. In addition, the SCMC controls the selection of LCBs in all zones. Once an LCB is selected in a zone, subscribers (i.e., smart devices) are moved from the previous LCB to the newly selected one.

\vspace{-5mm}

The SCMC is responsible for the following tasks:

\begin{itemize}
\item{Detection of zone changes of vehicles;}
\item{Computation of the estimated service time of vehicles;}
\item{Running of the proposed algorithm in the cloud.}
\end{itemize}

To detect a zone change in a vehicle, the SCMC computes the current zone number at time $t$ and compares it with the zone number at time $t-\delta$. Consequently, if the two values are different then the vehicle has entered a different zone at time $t$. The method for computing the zone number is later detailed in in Section~\ref{experiment_settings}.

When vehicle $v_i$ enters a zone, the SCMC computes the estimated service time $s_i$ then use the proposed algorithm to decide on the selection of $v_i$ as the LCB of the zone.

Now, if vehicle $v_j$ is working as an LCB in the zone and the proposed algorithm decides to select $v_i$ as the new LCB then the SCMC stops $v_j$ from working as LCB and move all subscribers to $v_i$. Switching between the LCBs results in a delay as subscribers are moved from the old LCB to the new one. Also, an incentive is used with every new LCB as explained earlier

The proposed algorithm utilizes an ensemble of TBO algorithms as explained in Section~\ref{heuristicSection}. Once a vehicle is selected as an LCB, all smart devices in the zone either publish their data or subscribe for services through the LCB.

\section{Online Hiring Algorithms}
\label{onlineHiringAlgorithms}

The problem of finding the best candidate is studied first in the classical online secretary problem where $M$ candidates is interviewed in random order with the goal of maximizing the probability of finding the best candidate. Later different variations of the secretary problem were proposed.

Recently, the problem is revisited and studied using hiring algorithms, which have been shown to be effective in selecting best candidates by many companies. In fact, instead of selecting employees manually, many companies like Google prefer to use hiring algorithms for automating this task in order to save time and efforts ~\cite{hiringPaper_2010,hiringPresentation_2007}. There are different strategies of the hiring algorithm such as TBO, `hire above minimum or maximum', and `Lake Wobegon'\footnote{Lake Wobegon refers to a fictional town conceived by the author Garrison Keillor, where ``\textit{all the women are strong, all the men are good looking, and all the children are above average}.''} (`hire above mean or median')~\cite{master_thesis_2010}.

In this paper, we investigate the use of a TBO algorithm using different threshold values. This algorithm is based on a fixed threshold $\tau$. In this algorithm, vehicle $i$ is selected only if its estimated service time $s_i$ is greater than $\tau$.

\section{Proposed Heuristic Solution}
\label{heuristicSection}

In this paper, we propose an algorithm that strives to maximize the total service time of a zone using vehicles as LCBs. The idea is simply to run an ensemble of $N$ TBO algorithms in passive mode while selecting only one of them to be active at any point in time. By passive, we mean an algorithm makes a decision for whether a given vehicle should be selected to serve as a data broker but the decision is not executed. This is done in order to collect performance metrics needed to compare the performance of the different algorithms in the ensemble.

The proposed algorithm is capable of analyzing the history of all TBO algorithms in its ensemble in terms of the service time. Additionally, to secure a best performance, the proposed algorithm may switch to another algorithm in its ensemble every $D$, where $D$ is a constant number of time units.
When the SCMC detects a zone change in a vehicle (i.e., the vehicle enters a different zone), it runs the proposed algorithm. The status of the proposed algorithm, that is maintained by the SCMC, comprises:

\begin{enumerate}
\item{\textit{Active algorithm};}
\item{\textit{Cumulative service time} for each of the $N$ algorithms;}
\item{\textit{Vehicle selected as the data broker} (if any);}
\item{\textit{Service time} of the selected vehicle (if any);}
\item{\textit{Remaining time until changing the active algorithm is explored}.}
\end{enumerate}

\begin{algorithm}[!th]
\caption{Algorithm for selecting the best LCB (in terms of the service time)}
\label{alg_proposed}
\begin{algorithmic}[1]
\STATEx \textbf{Input:} vehicle service time $S$.

\STATEx \textbf{Initialization:}
\STATE Set all algorithms as passive
\STATE Set active = select an algorithm number randomly.

\STATEx \textbf{Executed when zone change is detected:}
\FOR {each of the 9 algorithms}
\STATE Run the algorithm
\IF {Decision is accept}
\STATE Add $S$ to the cumulative service time.
\IF {this algorithm number = active}
\STATE Accept the vehicle
\ENDIF
\ELSIF {this algorithm number = active}
\STATE Reject the vehicle
\ENDIF
\ENDFOR

\STATEx \textbf{Executed every $D$ minutes:}
\STATE Set best = algorithm number that has the maximum cumulative service time
\IF {active $\neq$ best}
\STATE Set active = best
\ENDIF
\STATE Set cumulative service time for every algorithm to zero.
\STATEx \textbf{Output:} decision (accept or reject)
\end{algorithmic}
\end{algorithm}

\begin{figure}[!h]
\centering
\includegraphics[width=0.40\textwidth,trim={0 0 0 0},clip]{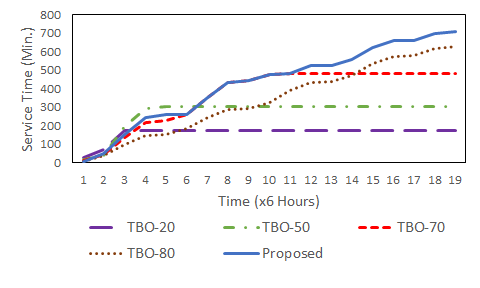}
\caption{Proposed algorithm switches to the best performing algorithm in terms of service time over time.}
\label{fig:overTime_withProposed}
\vspace{-3mm}
\end{figure}

Algorithm (\ref{alg_proposed}) shows the three parts of the proposed algorithm. In the first part, one of the algorithms in the ensemble is selected randomly to serve as the active algorithm and all other algorithms in the ensemble are set to passive. In the second part, all the algorithms in the ensemble are executed and the cumulative serve time is updated accordingly. Moreover, the decision made by the active algorithm is committed while decisions of other algorithms are ignored. The last part is only executed after $D$ time units had passed. Furthermore, the algorithm with the maximum cumulative service time is set as the active algorithm while setting the other algorithms as passive. In other words, every $D$ time units, the proposed algorithm compares all the algorithms in terms of cumulative service time and switches to another algorithm in case the current active algorithm is not the best. The switching behavior of the proposed algorithm is shown in Fig.~\ref{fig:overTime_withProposed}.

\section{Illustrative Example}

In this section, we discuss the proposed algorithm by utilizing two TBO algorithms (TBO-10 and TBO-20). We assume that at $t_0$, TBO-20 is randomly selected by the proposed algorithm (i.e., TBO-20 is active) while TBO-10 is initialized in passive mode. The performance in terms of average service time for the two algorithms is recorded every $D$ minutes (see Algorithm \ref{alg_proposed}) as shown in Table~\ref{table_example}. The proposed algorithm starts with a weak performance initially at time $t_1$, but then the performance enhances over time and by time $t_4$, the performance of the proposed algorithm surpasses that of the two TBO algorithms. In $t_1$, the proposed algorithm detects that TBO-10 is performing better and thus switches to TBO-10 (set TBO-10 as active and TBO-20 as passive). However, in $t_3$ the proposed algorithm switches back to TBO-20 since it starts performing better than TBO-10.

\begin{table}[!th]
	\centering
	\caption{Performance of the proposed algorithm compared with two TBO algorithms.}
	\label{table_example}
    \begin{tabular}{c}
        \includegraphics[width=.47\textwidth]{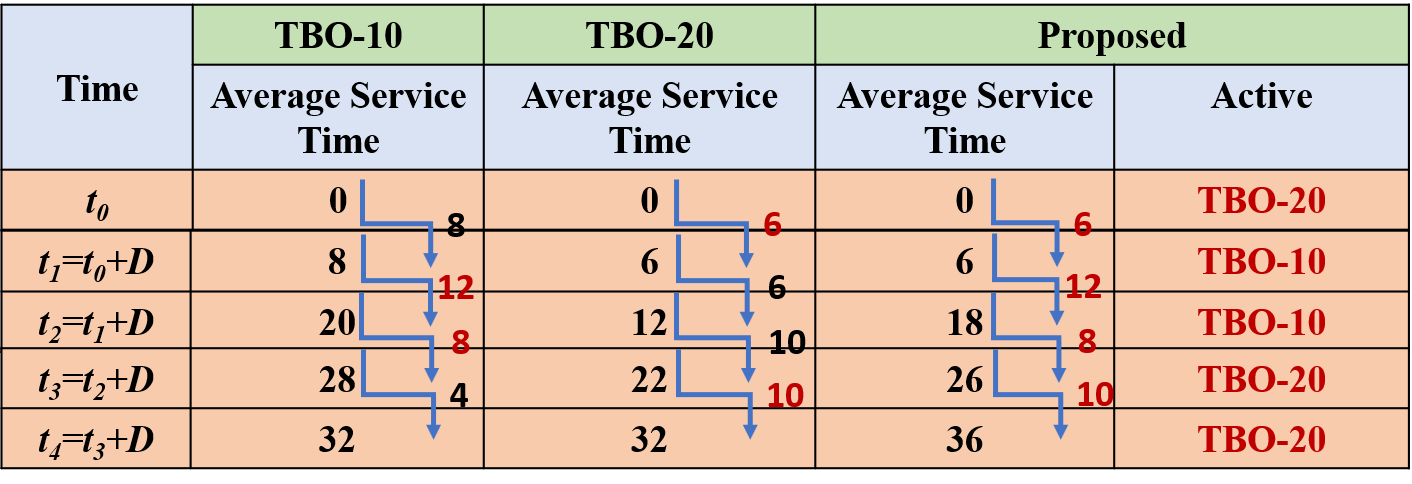}
    \end{tabular}
    \vspace{-5mm}
\end{table}

\section{Experimental Results}
\label{experiment_results}

In this section, we describe the dataset used in our experiment, explain the experiments' settings, and evaluate the performance of the proposed algorithm by comparing it with nine baseline `hire above a threshold' online algorithms using real vehicular traces of the Shanghai dataset. Finally, we discuss the results and present the insights learned from our experiments.

\subsection{Dataset and Experimental Settings}
\label{experiment_settings}

In this paper, we conducted our experiments using the Shanghai dataset, which consists of taxi traces observed in the city of Shanghai in China. To trace their positions, every taxi vehicle is equipped with a GPS unit. Additionally, each taxi has a GPRS wireless communication modem, which is used for sending GPS location along with other information to a data center. This dataset was collected in 2007 by monitoring 2,109 taxis. Moreover, the information sent by the taxis to the data center includes the taxi ID, the timestamp, the longitude and latitude, the speed, and the heading direction~\cite{shanghai_2007}.

To divide the city into zones, we encoded the longitude and latitude, which represents the geographical location, into a string of seven characters using the \texttt{GeoHashing} method~\cite{geoHashing_2015}. Every string produced using the GeoHashing method represents a zone in the city. Additionally, a string of seven characters encoding divides the globe into a number of zones, each of 153 by 153 meters, which is within the communication coverage of vehicles (i.e., RSU). Next, we convert the zone string to a number using hashing. Also, we filter the dataset by removing zones that have no traffic activity.

The used dataset is only one day long while the proposed algorithm needs more time to start working efficiently. Consequently, to have datasets with longer periods, we replicate the one-day dataset to create 5, 10, 15, 20, and 25 days datasets.

To study the effect of different traffic scenarios on the performance of the proposed algorithm, we divide the city based on the traffic volume into three areas: \textit{low}, \textit{medium}, and \textit{high traffic} areas. To compute the traffic volume per zone, the average number of vehicles per zone along with the standard deviation are computed. We noticed that the standard deviation is greater than the average. Therefore, based on $N_z$, the number of vehicles in zone $z$, each zone is categorized as follows:
\begin{itemize}
    \item \textit{Light traffic zone}: $N_z <$ average
    \item \textit{Medium traffic zone}: average $\leq N_z \leq$ standard deviation 
    \item \textit{High traffic zone}: $N_z >$ standard deviation
\end{itemize}

To test the capabilities of the proposed algorithm, we set the proposed algorithm to utilize nine TBO algorithms. We use the terminology \textit{TBO-x}---where $x$ is the threshold value as a percentile of the service times in a zone. For example, TBO-10 refers to the TBO algorithm with a threshold value of 10 percentile; the threshold value of TBO-20 is set to the value of the \nth{20} percentile of $L$, and so on for the other seven TBO algorithms. First, list $L$ is constructed from estimated service times of all coming vehicles in a zone. Then $L$ is sorted and threshold values of every TBO-x online algorithm is set as the $x$-th percentile of $L$. 

It can be noticed that threshold values of \nth{0} and \nth{100} percentiles are never used. We compare TBO-0 online algorithm with a threshold value of \nth{0} percentile and TBO-10, TBO-0 has less average service time per zone and 3 times more average vehicle selections per zone than TBO-10. Thus, the \nth{0} percentile threshold value is not worth to be considered. As for using the \nth{100} percentile threshold value, the results in terms of the average service time per zone is too small to be considered. Finally, we set $D$ to 6 hours in all of the experiments to be consistent.

\subsection{Results Discussion}

To evaluate the performance of the proposed algorithm compared with the nine TBO algorithms, we run the algorithms in different experiment. Also, we observed that TBO-10 online algorithm is performing better than the other eight TBO algorithms in terms of service time. Consequently, the proposed algorithm is compared with TBO-10 online algorithm in terms of the average service time per zone and the average number of selections per zone as indicated in Table~\ref{table_percService}.

\begin{table}[!th]
	\centering
	\caption{Performance of the Proposed Algorithm Compared to TBO-10 Online Algorithm In Areas of High Traffic Volume}
	\label{table_percService}
    \begin{tabular}{c}
     \includegraphics[width=.47\textwidth]{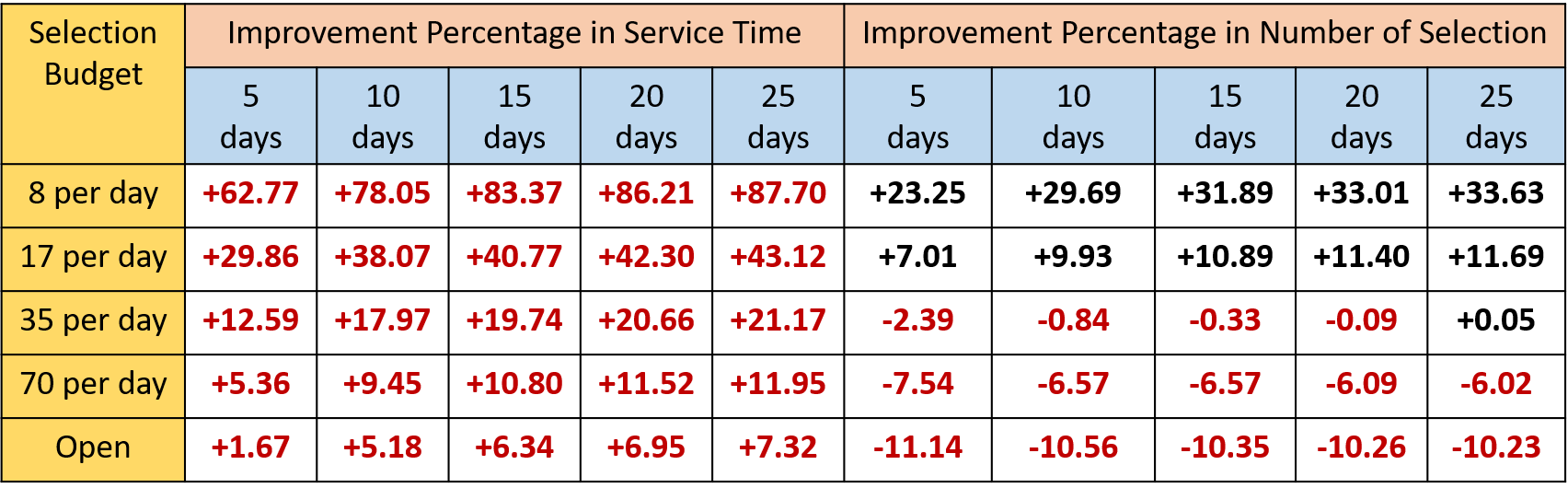}
    \end{tabular}
\end{table}

\begin{figure}[!th]
\centering

\begin{subfigure}[b]{0.40\textwidth}
\includegraphics[width=\textwidth]{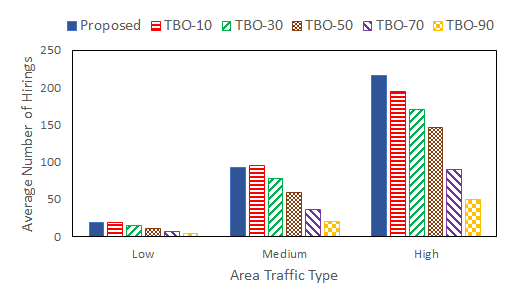}
\caption{Average number of selection per zone.}
\label{fig:5Days_hirings}
\end{subfigure}

\begin{subfigure}[b]{0.40\textwidth}
\includegraphics[width=\textwidth]{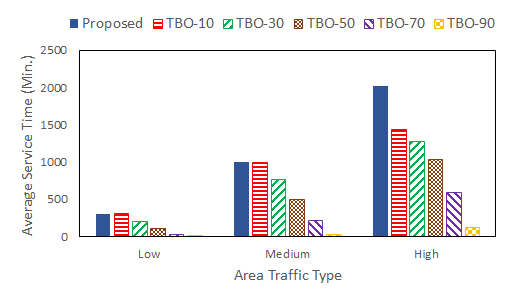}
\caption{Average service time per zone.}
\label{fig:5Days_service}
\end{subfigure}
\caption{Performance of algorithms for different traffic volumes for 5 days results. \textit{Proposed algorithm delivers more service time than other algorithms in high traffic area.}}
\label{fig:5Days}
\vspace{-5mm}
\end{figure}

\subsubsection{Measuring the performance of the proposed algorithms for different traffic volumes}

The city is divided based on traffic volume into three areas: low traffic area, medium traffic area, and high traffic area. Results are recorded in terms of the average service time per zone and the average number of selections per zone.

We found that the proposed algorithm is not as competitive in low and medium traffic areas as in high traffic area as shown in Fig.~\ref{fig:5Days}. This is because the areas of low and medium traffic have fewer vehicles and longer periods with no vehicles. Therefore, we focus in our discussion on high traffic area results.

\subsubsection{Measuring the performance of the proposed algorithms for different selection budgets and time intervals}

We test the algorithms for a different number of days using different selection budgets. Selection budgets are set based on different fractions of the average number of vehicles per zone, which is 35 for the used dataset. In each experiment, results are recorded per city area.  

Table~\ref{table_percService} shows that the proposed algorithm performs better in terms of average service time in all selection budget settings. Also, the proposed algorithm collects more profit in terms of service time and reduces the cost in terms of the number of selections in scenarios with more selection budget. Moreover, the proposed algorithm collects more profit in scenarios with limited selection budget at the cost of enhanced switching between the vehicular data brokers. As more selection budget is available, the proposed algorithm becomes more inclined towards optimizing cost, which provides a balance between cost and profit.

\section{Conclusions and Future Work}

In this paper, the problem of selecting vehicles to serve a data broker in support of smart community applications is considered. The selection process strives to achieve the maximum service time. An algorithm is proposed that utilizes an ensemble of TBO online algorithms by running them altogether in passive mode and selecting the one that performs best in the recent history. The proposed algorithm is evaluated using real taxi traces from the city of Shanghai in China and compared against a baseline of nine TBO algorithms. Experiments with these traces demonstrate that the proposed algorithm outperforms TBO algorithms presented in the literature in high traffic volume regardless of the selection budget by performing better on the service time. Also, the proposed algorithm reduces the number of selections with high selection budgets.
In the future, we plan to evaluate the proposed algorithm analytically to provide performance guarantees, in terms of competitive ratio, in worst-case scenarios.

\ifCLASSOPTIONcaptionsoff
  \newpage
\fi

\bibliographystyle{IEEEtran}
\bibliography{biblo}

\end{document}